# An innovative materials design protocol for the development of novel refractory high-entropy alloys for extreme environments

O. El Atwani[1*], H.T. Vo[1], M. Tunes[1], C. Lee[2], A. Alvarado[3,4], N. Krienke[5], J.D. Poplawsky[6], A.A. Kohnert[1], J. Gigax[2], W.-Y. Chen[7], M. Li[7], Y. Wang[1], J.S. Wróbel[8], Duc Nguyen-Manh[9,10], J.K.S. Baldwin[2], U. Tukac[11], E. Aydogan[11], S. Fensin[2], E. Martinez[4]

[1]Materials Science and Technology Division, Los Alamos National Laboratory, Los Alamos, NM, United States

[2] Center for Integrated Nanotechnology, Los Alamos National Laboratory, Los Alamos, NM

[3] Theoretical Division, Los Alamos National Laboratory, Los Alamos, NM, USA

[4] Departments of Mechanical Engineering and Materials Science and Engineering, Clemson University, Clemson, SC, USA.

[5]Materials Science and Engineering, University of Wisconsin-Madison, Madison, WI, United States

[6]Materials Science and Technology Division, Oak Ridge National Laboratory, Oak Ridge, TN, United States

[7]Division of Nuclear Engineering, Argonne National Laboratory, Lemon, IL, United States

[8] Faculty of Materials Science and Engineering, Warsaw University of Technology, ul. Wołoska, 02-507 Warsaw, Poland

[9] Culham Center for Fusion Energy, United Kingdom Atomic Energy Authority, Abingdon OX14 3DB, UK

[10] Department of Materials, University of Oxford, Oxford OX1 3PH, UK

[11] Metallurgical and Materials Engineering, Middle East Technical University, Ankara, Turkey

[*] Corresponding author: Osman El-Atwani (osman@lanl.gov)

## Abstract

In the quest of new materials that can withstand severe irradiation and mechanical extremes for advanced applications (*e.g.* fission reactors, fusion devices, space applications, etc), design, prediction and control of advanced materials beyond current material designs become a paramount goal. Here, though a combined experimental and simulation methodology, the design of a new nanocrystalline refractory high entropy alloy (RHEA) system is established. Compositions of this alloy, assessed under extreme environments and *in situ* electron-microscopy, revealed both high mechanical strength and thermal stability, grain refinement under heavy ion irradiation and outstanding irradiation resistance to dual-beam irradiation and helium implantation, marked by remarkable resistance to defect generation, growth and coalescence. The experimental and modeling results – which demonstrated notable agreement – can be applied to design and rapidly assess other alloys subjected to extreme environmental conditions.

1. **Introduction**

Clean energy production is the cornerstone of our time. Options for sustainable clean energy include advanced power generation systems that have the potential to drastically reduce the emission of greenhouse gases. These advanced systems are often required to operate under harsh conditions to optimize efficiency, which poses several challenges for the available materials. An example of advanced power system is one that is associated with fusion energy.[1] Beyond traditional fission-based systems, fusion reactors not only promise nearly unlimited clean energy, but also avoid the generation of long-life radioactive waste produced in fission devices. One remaining challenge is that of materials, which can withstand extreme conditions of radiation, temperature, and stress, with long-term steady properties for the power plant to be economically viable.[2-4] A key component in current tokamak designs is the divertor, which will be in contact with the deuterium-tritium (D-T) plasma and sustain severe fluxes of particles (helium (He) ash, D and T) and heat, along with radiation damage induced by high-energy neutrons.[2,3] Tungsten (W) is the current element of choice for the plasma-facing components (PFCs) due to its beneficial properties in terms of heat conduction, mechanical response, and T retention.[5-7] However, He bubble formation, surface morphology evolution and neutron damage compromise its ability to reach the viability requirements.[8-14] Several strategies have been proposed to enhance the properties of the material facing the plasma. Reducing the grain size, hence increasing the density of interfaces, is one of them.[15] Grain boundaries are known to promote defect annihilation and therefore, decrease the overall amount of defects generated by irradiation leading to deleterious effects on the material properties.[16,17] However, this approach can suffer from some drawbacks in pure materials such as the thermal instability of the nanocrystalline grains (coarsening at the application temperature).[18,19] Another approach is to develop alloys where elements can increase strength, act as defect annihilation and recombination sites [20] and enhance thermal stability of the material. Recently, a novel set of alloys based on equiatomic compositions of several principal elements (multi-principal elements alloys (MPEAs) or high-entropy alloys (HEA)) have been developed. [21-23] The configurational entropy of mixing in multicomponent alloys tends to be the major thermodynamic driving force to stabilize the solid solution based on simple underlying face-centered cubic (FCC) or body-centered cubic (BCC) crystalline structures.[24] Equiatomic compositions maximize the entropic term of the Gibbs free energy of mixing, promoting the formation of random solutions versus intermetallic phases or phase decomposition. [25]

W-based refractory HEAs (RHEAs) have been recently developed in the context of high-temperature applications, showing high melting temperature (above 2873 K) and superior mechanical strength at high-temperatures compared to Ni-based superalloys or pure W. [26,27] Combining the two approaches above, the authors have recently developed a refractory low-activation HEA based on W-Ta-Cr-V.[28] Its response to loop formation under ion irradiation [28] and He implantation [29] is superior compared to previously developed W systems, showing no noticeable dislocation loop formation and smaller He bubbles with no radiation-induced segregation at grain boundaries upon heavy-ion irradiation and He implantation, respectively.

Additionally, this quaternary HEA also has a hardness of ~ 14 GPa, although with reduced ductility and toughness at low temperatures. However, this material demonstrated Cr- and V-rich precipitates which could be detrimental to mechanical properties in terms of embrittlement.

In this study, we have developed a design strategy to further improve the overall response of the W-Ta-Cr-V RHEA. The aim of the design is to develop a material which higher irradiation resistance, high thermal stability, enhanced control over the morphology and no precipitation at reactor-relevant temperatures while following the criteria for enhanced ductility suggested in literature.[30] It has been shown that BCC RHEAs can show enhanced intrinsic ductility with elongation between 6 and 15% if the valence electron concentration (VEC) remains below 4.4. [31-33] Lowering the VEC below 4.4 has also been shown to change the failure mode from screw dislocation glide to shear deformation, and this was considered as a first strategy to discover HEAs with better ductility. [30] Another option to increase the ductility was to increase the VEC number to over 6.87 which was suggested to induce an FCC phase to the already existing brittle BCC phase when the VEC is between 6.87 and 8 or to form a single stabilized FCC phase when the VEC is over 8, and this was considered as a second strategy to enhance ductility in HEAs.[34] In addition, the ability of the material to form a single phase was also shown to be dependent on the enthalpy of mixing ($\Delta H_{mix}$), atomic size mismatch ($\delta$) and the omega parameter ($\Omega$) as defined below in the results section. The value for $\Delta H_{mix}$ were recommended be in the range of -15 to -5 meV per atom, while $\delta$ should be smaller than 6.6 and $\Omega$ larger than 1.1.[35-37] Based on these strategies, addition of group IV elements (Ti, Zr, Hf) to W-based alloys can improve intrinsic ductility, while maintaining a lower VEC number. As the material will need to withstand extremely high temperatures, we have chosen Hf as the additional element, as it has the highest melting temperature among the elements in group IV.

In the present work, results from developing such a material system manufactured via magnetron sputtering deposition with nanocrystalline grains will be discussed. This work describes in detail an innovative RHEA design protocol, including insights from modeling strategies and experimental methodologies. Computational thermodynamics approaches such as the CALculation of PHAse Diagrams (CALPHAD) method and density-functional theory informed cluster expansion formalisms have been used to select optimal compositions and predict thermodynamic properties. These compositions were then manufactured experimentally to test its ion irradiation and hardness response and validate the modeling predictions. Although the design aimed at following the criteria suggested for ductile HEAs, the irradiation response, thermal stability, strength and morphology predictions are being elucidated here. The ductility enhancement can be examined (outside the scope of this paper) when the designed material is produced in bulk and large grains form to examine the intrinsic ductility of the materials and avoid the loss in ductility in the nanocrystalline grain regime.[38] The results show that the alloy has remarkable microstructural stability along with promising ion irradiation response upon single and dual-beam ion irradiation conditions. This outstanding material response can be attributed to a combination of factors, including high density of stable grain boundaries, even showing grain refinement, chemical complexity altering defect recombination rates, and a decrease in the order-disorder transition temperature (ODTT) as compared to the original four element RHEA. The established design protocol can be further utilized to design and synthesize new RHEAs and constitutes a material design paradigm with high throughput morphology predictions.

## 2. Methodologies

### 2.1. Fabrication of the W-Ta-Hf-Cr-V alloy

The material was synthesized using magnetron-sputtering deposition from metal targets of 99.99% purity using power of 200, 400, 50, 350, 25 Watts for W, Ta, Cr, V, and Hf, respectively. The deposition was performed at room temperature and 3 mTorr pressure at no bias voltage. Two sets of depositions were performed: (1) 100 nm thin film on NaCl substrate and (2) 3 µm film on W substrate. Transmission electron microscopy (TEM) samples were then prepared by floating the film on a standard molybdenum TEM grid using 1:1 ethanol/water solution. Nanoindentation was performed on the 3 µm film on W substrate. The thin (for in-situ TEM) and the thick films (for mechanical properties and atomic probe tomography, APT, studies) had compositions $W_{29.4}Ta_{42}Cr_{5.0}V_{16.1}Hf_{7.5}$ and $W_{31}Ta_{34}Cr_{5.0}V_{27}Hf_{3.0}$, respectively.

### 2.2. Pre-irradiation characterization of the thin film TEM samples

Before ion irradiation, the films were analyzed using energy dispersive X-ray (EDX) spectroscopy and selected area diffraction (SAED) in an FEI Titan 80-300 TEM operated at 300 keV. The material was then annealed *in situ* within the TEM at 1173K for 10 minutes. EDX and SAED were performed on the thin film sample in the as-deposited and post annealed, implanted and irradiated conditions.

### 2.3. In-situ TEM ion irradiation of the thin film samples

The RHEA material was irradiated *in situ* at the Intermediate Voltage Electron Microscope (IVEM)-Tandem Facility at Argonne National Laboratory with 1-MeV $Kr^{+2}$ and 16 keV $He^+$. Two irradiation conditions were performed: (1) dual beam irradiation with 1-MeV $Kr^{+2}$ and 16 keV $He^+$ and (2) single beam implantation with 16 keV $He^+$. In the dual beam ion irradiations, the material was irradiated to 8.5 displacements per atom (dpa) and ~9 % He. The implanted He per dpa ratio was ~1.07 % He/dpa. The corresponding Kr and He fluences were $2.74 \times 10^{15}$ and $5.84 \times 10^{16}$ ions/cm$^2$ with the average fluxes of $4.86 \times 10^{11}$ and $1.04 \times 10^{13}$ ions/cm$^2$/s, respectively. In the single beam implantation, the material was implanted with He up to ~ 9 %, same amount as in the dual-beam irradiations. The dpa and He implantation profiles were calculated using the Kinchin-Pease model in the Stopping & Range of Ions in Matter (SRIM) Monte Carlo computer simulation code (version 2013)[39] and 40 eV [40] was taken as the displacement threshold energy for all elements. The dpa and He implantation profiles are presented in the supplementary material. The irradiation temperature in both dual and single beam irradiation experiments was set to 1173 K. *In-situ* videos were collected for irradiation-induced damage quantification. The damage quantification procedure is well described in ref. [41] The overall change in volume was found using $\Delta v/v = \frac{4}{3}\pi r_c^3 N_v$ where $N_v$ is the bubble density in a 100 nm thick foil and $r_c$ is the radius of the bubble.

### *2.4. Ex situ ion irradiation of the thick film samples*

The *ex-situ* ion irradiations were performed with 400 keV Ar$^{+2}$ ions on a 200 kV Danfysik Research Ion Implanter at the Ion Beam Materials Laboratory at Los Alamos National Laboratory. The beam flux and the fluence were $1.8 \times 10^{12}$ ions/cm$^2$/s and $8.2 \times 10^{15}$ ions/cm$^2$ (10 dpa), respectively. The sample was mounted on a heating stage with silver paste and the stage temperature was kept at 1073 K during the irradiation and monitored continuously with a thermocouple mechanically attached to the heating stage. The *ex-situ* irradiation beam parameters were chosen to match the damage process of the *in-situ* irradiation conditions at IVEM for subsequent atom probe tomography (APT) analysis. The dpa and He implantation profiles are presented in the supplementary material.

## 2.5. Post-irradiation characterization of the thin film samples

SAED and high-resolution EDX measurements were performed after the irradiation. APT was performed on the as-received, after-annealing, and after-irradiation conditions. CAMECA's Integrated Visualization and Analysis Software (IVAS) was utilized to reconstruct and analyze the APT data. APT samples were fabricated using standard lift out and sharpening methods as described by Thompson *et al.*[42] Briefly, wedges were lifted out, mounted on Si microtip array posts, sharpened using a 30 kV Ga+ ion beam, and cleaned using a 2 kV Ga+ ion beam. For the irradiated samples, the top of the needle was located as close to the surface as possible (< 50 nm from the surface). The APT experiment was run using a CAMECA LEAP 4000XHR in laser mode with a 30 K base temperature, 80 - 100 pJ laser energy, a 0.5 % detection rate, and a pulse repetition rate set to capture all elements in the mass spectra.

## 2.6. Mechanical properties assessment of the thick film samples

Nanoindentation tests were performed on the as-deposited and post-irradiated materials using a Keysight G200 Nanoindenter with a diamond, pyramidal (Berkovich) tip to a final displacement of 1000 nm with a constant strain rate (loading rate divided by the load) of 0.05 $s^{-1}$. Continuous stiffness measurements (CSM) were performed at a frequency of 45 Hz and 2 nm displacement amplitude. Since the irradiated layer is 200 - 300 nm and size effect can be around three times the indented depth (for brittle materials), the hardness and modulus measurements were obtained at 200 nm indentation depth.

## 2.7. Modeling techniques

### Density functional theory

Density functional theory (DFT) calculations were performed using the Vienna Ab Initio Simulation Package (VASP).[43] For the exchange and correlation, a generalized gradient approximation of the Perdew-Burke-Ernzerhof form (GGA-PBE), with projector augmented wave method was used. Both the semi-core p electron and magnetism were not included in this study since their contribution does not significantly affect the calculations. The convergence criteria for energy were set to $10^{-5}$ eV per cell. The cells were relaxed to a force convergence criterion of $10^{-3}$ eV Å$^{-1}$. The Monkhorst-Pack mesh spacing[44] was such that it corresponded to a $14 \times 14 \times 14$ k-point mesh of a two-atom BCC cell. The plane-wave cutoff energy used was 400 eV.

The Alloy Theoretic Automated Toolkit (ATAT) package[45,46] was used to generate a cluster expansion (CE)-based configurational energy expression fitted to DFT energies. A modified database of 58 initial structures for each binary subsystem was used following D. Nguyen-Manh *et al*.[47] For ternary subsystems, 94 ternary structures were constructed from initial binary structures by replacing the atoms in one of the nonequivalent positions from the symmetry point of view with the third type of atom.[48] For quaternary and quinary structures, a database from ref [49] was used and modified for the W-Ta-Cr-V-Hf system. In total, 1511 BCC structures including quinary, binaries, ternaries, and quaternaries subsystems of the W-Ta-Cr-V-Hf HEA were used in the fitting. During DFT relaxation, the volume and shape of the cell were allowed to change. Only structures without large distortions compared to the starting configuration were used in the fitting. From ATAT's toolkit package, the *checkrelax* function was used to ensure the square-root sum of each element of the strain tensor squared no larger than 0.1. Additionally, the common neighbor analysis algorithm in OVITO [49] was used to ensure the structure remained mainly in a BCC phase.

*Cluster Expansion Formalism*

From DFT, the enthalpy of mixing can be computed as:

$$\Delta E_f^{DFT} = \frac{E_{DFT} - \sum_{m=1}^{n} N_m E_m^{ref}}{N} \qquad \text{Eq. (1)}$$

where $E_{DFT}$ is the energy of the system as calculated from *ab initio*, $N$ is the total number of atoms in the supercell, $n$ is number of components in the alloy, $N_m$ is the number of atoms of type *m*, and $E_m^{ref}$ is the reference energy of atom type *m*. The reference energies, $E_m^{ref}$ were calculated using the DFT methodology described above with values of -9.51075, -9.77225, -11.8619, -8.94221, and -13.0112 eV per atom for Cr, Hf, Ta, V, and W, respectively, in a BCC crystalline structure.

In the cluster expansion formalism, the enthalpy of mixing can be expressed using an Ising-like Hamiltonian, [46]

$$\Delta E_f^{CE} = \sum_{\omega} m_\omega J_\omega \langle \Gamma_{\omega\prime}(\vec{\sigma}) \rangle_\omega \qquad \text{Eq. (2)}$$

where $\vec{\sigma}$ specifies an atomic configuration in the form of configuration variables. The summation is performed over all clusters $\omega$. The clusters $\omega$ are distinct under symmetry operations of an underlying lattice, the number of equivalent clusters is obtained by multiplicity $m_\omega$. $J_\omega$ are the concentration-independent effective cluster interactions (ECIs). $\langle \Gamma_{\omega\prime}(\vec{\sigma}) \rangle$ are point function products of occupational variables on averaged cluster $\omega'$.

The effective cluster interactions were computed from first principles through a structural inversion method (SIM).[50] Through SIM, one can utilize the energy corresponding to a relaxed set of structures, via DFT, to calculate the cluster functions, create a set of linear equations, and fit the ECIs. To determine the accuracy of the CE model a cross validation (CV) score is used. CV is the square root mean difference between *ab initio* energies to those obtained by CE:

$$CV = \sqrt{\frac{1}{n}\sum_{i=1}^{n}\left(\Delta E_{f,i}^{DFT} - \Delta E_{f,i}^{CE}\right)^2} \qquad \text{Eq. (3)}$$

where $\Delta E_{f,i}^{DFT}$ is the energy of structure *i* as calculated by DFT, and $\Delta E_{f,i}^{CE}$ the energy of the same structure predicted using CE from a least-square fit to the other (*n*-1) structural energies.

The Warren-Cowley short-range order (SRO) parameters are used to quantify the chemical ordering between pairs of different species up to the second nearest neighbor by:

$$\alpha_n^{ij} = 1 - \frac{y_n^{ij}}{c_i c_j}$$

Where $\alpha_n^{ij}$ is the chemical short-range order parameter, $y_n^{ij}$ is the probability of two atoms within n[th] neighbor shell and $c_i$ and $c_j$ are the concentrations of species *i* and *j*, respectively. $y_n^{ij}$ is calculated through a matrix inversion of correlation functions obtained via Canonical Monte Carlo (CMC). [51]

## 3. Results

The design stage of the new high entropy alloy started with the calculation of the thermophysical parameters, mainly enthalpy of mixing ($\Delta H_{mix}$), atomic size difference ($\delta$), and the omega parameter ($\Omega$) for several compositions. $\Delta H_{mix}$, $\delta$ and $\Omega$ were calculated using Equations 4, 5 and 6, respectively. $\delta$ and $\Omega$ were obtained directly from our DFT results. Other theoretical parameters such as density ($\rho$), melting temperature ($T_m$), and valence electron concentration (VEC) were calculated using the rule of mixture, as shown in Equation 7, where *i* represents the element type, $X$ represents the property, $r$ is the atomic radius, and $c$ represents the concentration of element $i$.

$$\Delta H_{mix} = \sum_{i<j} 4 H_{ij}^{mix} c_i c_j \qquad \text{Eq. 4}$$

$$\delta = 100 \, x \sqrt{\sum_{i=1}^{N} c_i (1 - r_i/r)^2} \qquad \text{Eq. 5}$$

$$\Omega = T_m \, \Delta S_{mix} \, / \, |\Delta H_{mix}| \qquad \text{Eq. 6}$$

$$X = \sum c_i \, (X)_i \qquad \text{Eq. 7}$$

The atomic radii and enthalpy used in the calculations are taken from ref.[52] and ref. [53] respectively. The base material for this design was the W-Ta-Cr-V HEA which was shown to possess high mechanical strength and radiation resistance to loop and He bubble formation under ion irradiation .[28] We first explored replacing V with Fe. This was done because it has previously been employed as an element of low radioactivity (for nuclear application requiring low activation materials). The calculations are shown in the supplementary material. However, the minimum achievable VEC after substituting Fe for V, and keeping the Fe concentration below 35% to maximize configurational entropy, was estimated to be 5.95 (very high to explore the first strategy) and the maximum was 6.55 (low to explore the second strategy) using CALPHAD, as implemented in the ThermoCalc software (with TCHEA5[54] and MOBHEA2 [55] databases). Furthermore, addition of Fe was shown to promote the formation of several intermetallic phases instead of single-phase solid-solution. Fe concentration was then further increased to increase the VEC and explore the second strategy (increase VEC to over 6.87). Three compositions were identified and the VEC was maximized by minimizing Ta and maximizing Fe. Lower melting points and the formation of intermetallic phases were highlighted as potential issues with these compositions. Due to failure of these approaches, a quinary HEA was planned to minimize VEC (first strategy). Group IV elements can then be used to pursue this approach (e.g. Ti, Zr and Hf). Hf was first selected with a minimal concentration to avoid a material with high activation rates when under in-reactor conditions and to maximize the probability of obtaining a single-phase HEA (all calculations including CALPHAD for selected compositions are in the supplementary material). In this case, 5.2 was shown to be the minimum VEC and certain compositions were shown to favor the formation of a single-phase BCC HEA.

As mentioned in the methodology section, two films were prepared for study in this investigation: 1) 100 nm thin film to study the thermal stability and the irradiation response to He implantation and dual beam irradiation, and 2) a 3µm film for morphology and hardness analysis via APT and nanoindentation respectively. The deposited thin and thick films with Hf had compositions of $W_{29.4}Ta_{42}Cr_{5.0}V_{16.1}Hf_{7.5}$ and $W_{31}Ta_{34}Cr_{5.0}V_{27}Hf_{3.0}$ which are predicted by CALPHAD to form a single-phase BCC structure over a wide temperature range (illustrated in the supplemental). We acknowledge that both produced films have different compositions which is due the challenge in depositing specific compositions from five different targets within this quinary system. However, it should be noted that different compositions, tested in thin film forms, led to similar single-phase BCC microstructures with remarkably similar irradiation responses (in terms of dislocation loop and cavity damage as well as grain size stability), and for clarification, only one composition is discussed throughout this paper. Hence, the thick film composition is expected to follow similar irradiation resistance. However, it is important to note that simulations performed on the thin film and thick film compositions revealed interesting results regarding the morphology and chemical segregation as discussed below. The morphology of the as-deposited, and after *in-situ* TEM annealing of the thin film to 1173 K are shown in Figures 1a and 1b, respectively. A single-phase HEA is clearly shown in the corresponding diffraction patterns. Two types of ion irradiations were then performed: 1) 16 keV He implantation to compare with other W-based materials and to maximize swelling and 2) dual beam (1 MeV $Kr^+$ + 16 keV $He^+$) to test the materials under reactor relevant conditions where heavy ions mimic fast neutron damage, albeit with a different energy spectrum, and the implanted He mimics the gas production from transmutation reactions. Previous work by El Atwani *et al.* on W under sequential and simultaneous dual beam conditions has shown that simultaneous dual beam leads to different loop and cavity damage evolution, compared to damage under single or sequential beams.[56] The amount of He/dpa, however, was kept very high in an effort to evaluate the quinary RHEA response to extreme conditions.

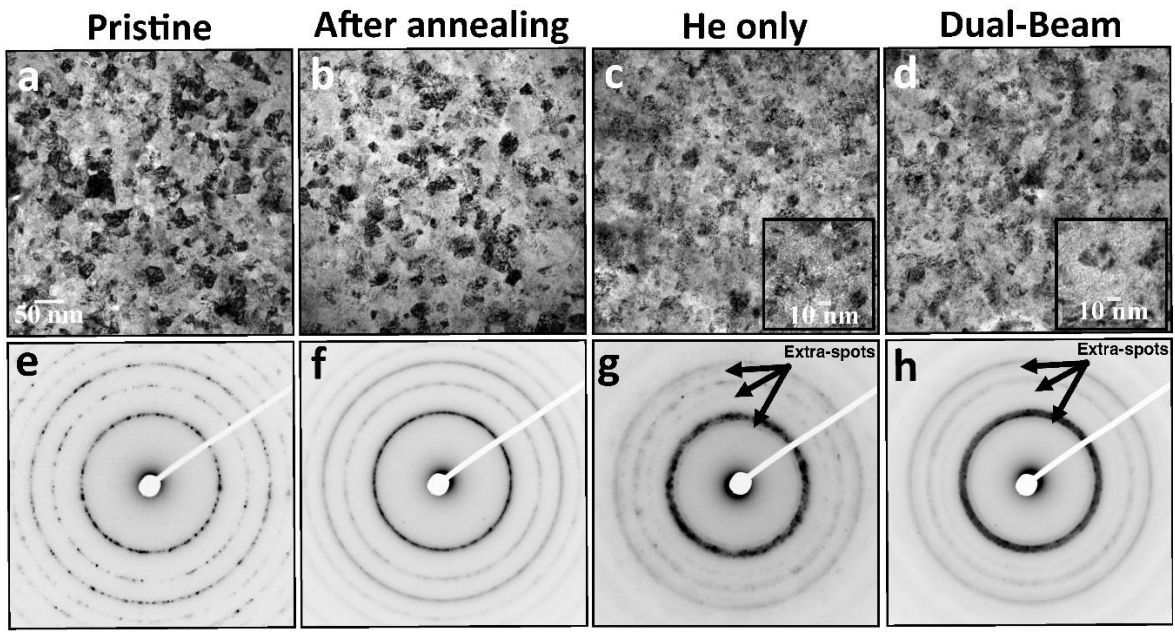

**Figure 1:** Bright field TEM (BF-TEM) images showing the nanocrystalline microstructure of the (a) as-deposited, (b) annealed, (c) single-beam He implanted, and (d) dual-beam irradiated $W_{29.4}Ta_{42}Cr_{5.0}V_{16.1}Hf_{7.5}$ 100 nm film HEA. The corresponding diffraction patterns are shown in (e)-(h).

TEM micrographs from post implantation and after dual beam ion irradiation samples are shown in Figure 1c and 1d, respectively. Thickening of the BCC rings as shown in Figures 1g and 1h indicates the presence of lattice strain, possibly due to defect formation and concentration gradients. It is also important to note that additional rings are present after irradiation. These rings were not identified, but they are anticipated to correspond to a different phase (*e.g.* shallow surface oxides due to irradiation). Experimental and modelling results (discussed below) indicate no metallic segregations in the grain matrices. Figure 2 shows the corresponding chemical distribution for the samples. Careful analysis of the elemental maps revealed that after annealing, Hf segregates to grain boundaries and V and Cr are also inhomogeneously distributed throughout the sample, thus indicating segregation. After irradiation, Hf starts to deplete from the grain boundaries (Hf concentration decreases compared to after annealed sample).

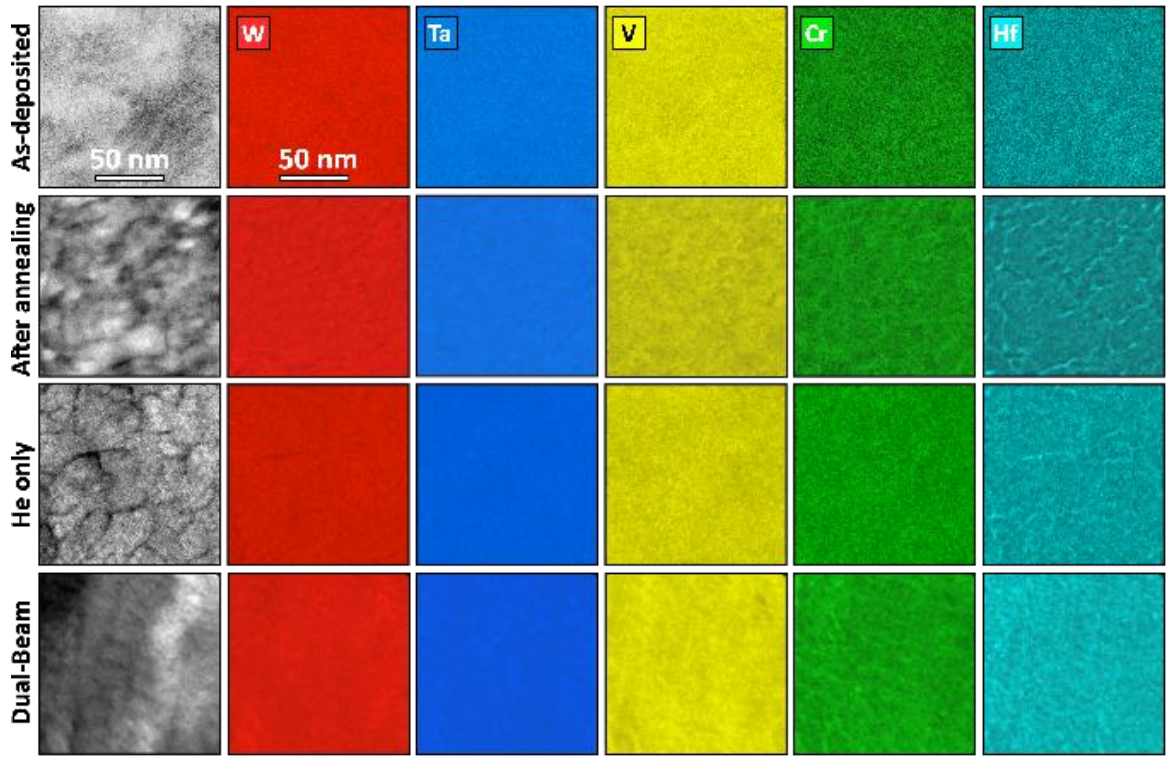

**Figure 2:** EDX chemical comparison via TEM of the as-deposited, annealed, He implanted, and dual beam irradiated $W_{29.4}Ta_{42}Cr_{5.0}V_{16.1}Hf_{7.5}$ 100 nm film HEA.

The thermal stability in terms of grain-size is studied via *in-situ* TEM annealing and irradiation and the results are plotted in Figure 3. The average grain size of the as-deposited sample was 24.5±1.3 nm. After annealing to 1173 K for ~ 30 minutes in the TEM, the grain size increased to 26.3±1.1 nm. After He implantation, the grain size further increased to 36.3±1.8 nm. However, during the dual beam irradiation, grain refinement occurred, and the average grain size dropped to 21.1±1.4 nm after a total irradiation time of ~94 minutes. The overall material stability in terms of phases and segregation is discussed further below.

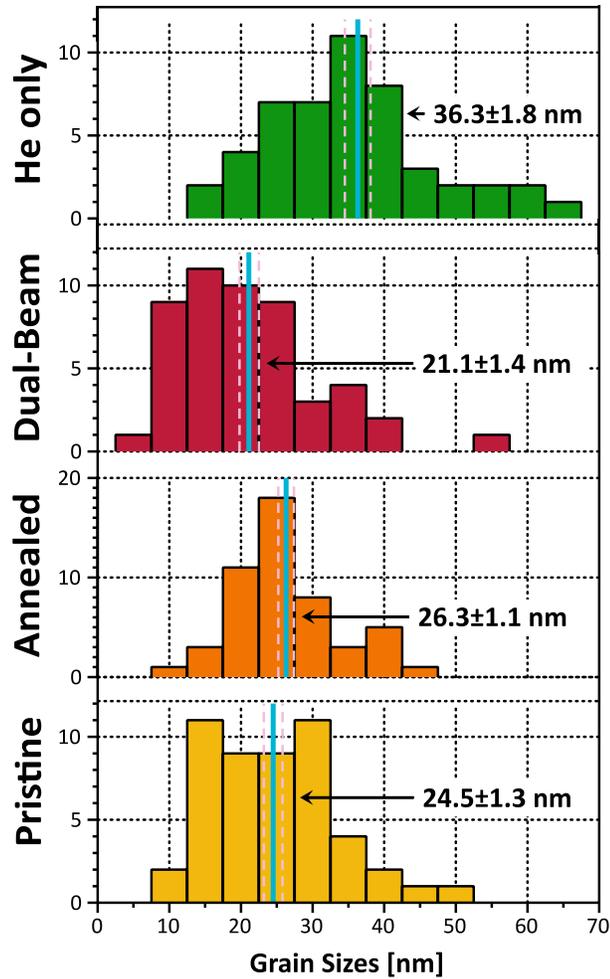

**Figure 3:** Quantification via BF-TEM micrographs of the grain-size distribution of the as-deposited, annealed, dual beam irradiated, and He implanted $W_{29.4}Ta_{42}Cr_{5.0}V_{16.1}Hf_{7.5}$ 100 nm alloy. Note: Displayed errors are the standard error of mean.

The irradiation resistance in the alloy is studied in terms of both dislocation loop formation and cavity formation. Dislocation loops were not detected even after 8.5 dpa and at 9.13% He implantation during the dual beam experiment (irradiation video is attached to the supplemental). Only cavities were observed. Quantification of the cavity density, average cavity size and total change in sample volume is shown in Figure 4. Cavities were visible in the microscope after 5 dpa irradiation in the case of dual beam and ~8.5% He implantation. No increasing trends in cavity volume, density or change in volume occur after that, and a saturation is evident for these dpa/He percentage ranges.

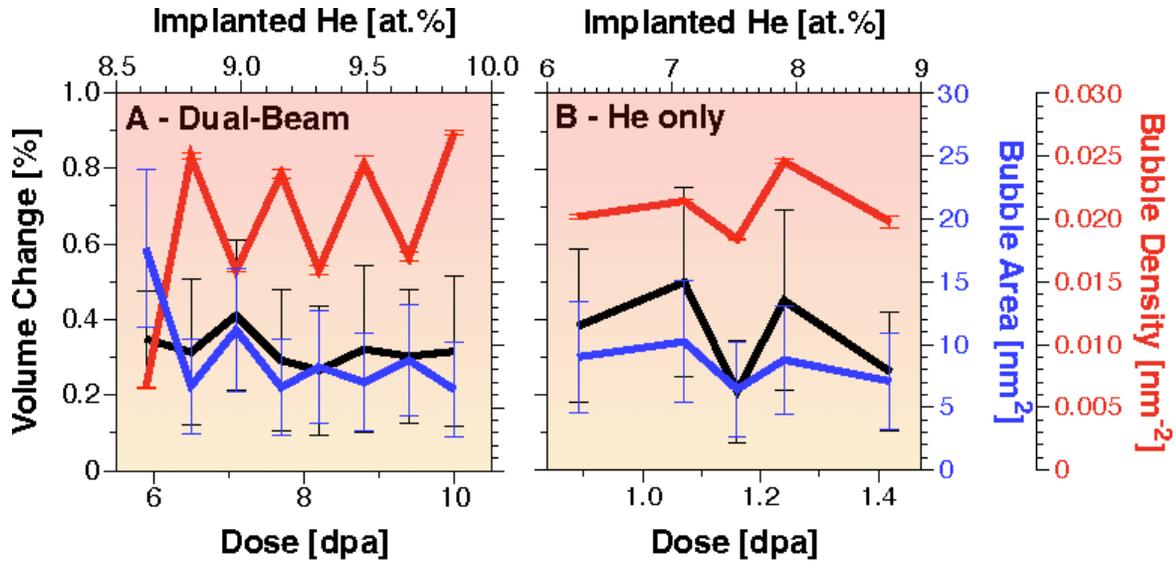

**Figure 4:** Plots showing the average volume change, average bubble area, average areal number density of the dual-beam irradiated (left) and He implanted alloy (right) as a function of dpa and percentage of implanted He in the $W_{29.4}Ta_{42}Cr_{5.0}V_{16.1}Hf_{7.5}$ 100 nm alloy.

To estimate the mechanical strength of a specific composition of this alloy system, nanoindentation was performed before and after annealing and after irradiation on the thick deposited specimen. This specimen had a composition of $W_{31}Ta_{34}Cr_{5.0}V_{27}Hf_{3.0}$ and was irradiated with 400 keV $Ar^{+2}$ ions at 1073 K to 10 dpa. The results are shown in Figure 5. The average hardness of the unirradiated sample was measured to be 13.25 GPa. After annealing the sample hardness increased to 16.25 GPa, while after irradiation the sample hardness reached a value of 20 GPa. The unirradiated sample is about 75% harder than pure W.[57] The change in hardness after irradiation will be discussed based on morphology and further analysis performed and discussed below.

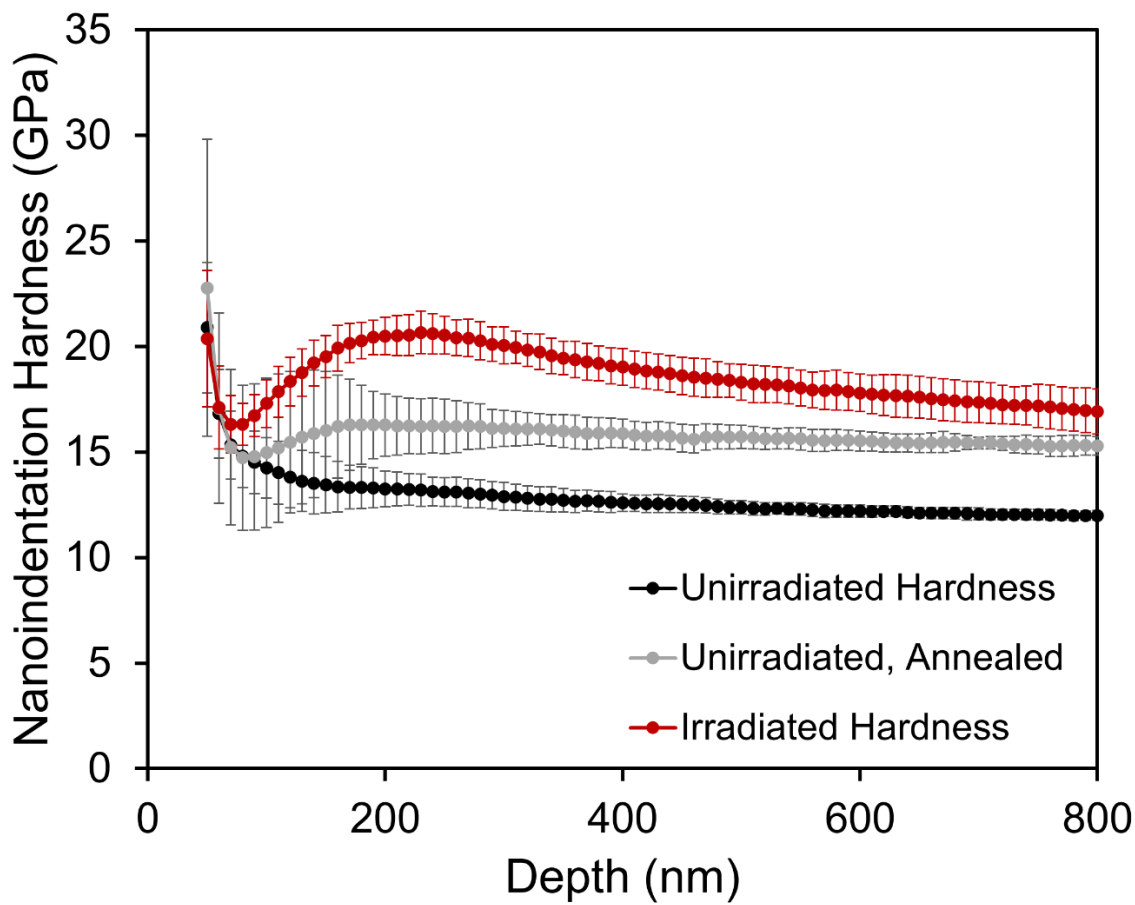

**Figure 5:** Nanoindentation results showing hardness vs depth of the $W_{31}Ta_{34}Cr_{5.0}V_{27}Hf_{3.0}$ thick HEA film before and after annealing, and post irradiation to 10 dpa.

To predict the morphology of the samples, first-principal calculations of phase stability and chemical short-range ordering in the $W_{29.4}Ta_{42}Cr_{5.0}V_{16.1}Hf_{7.5}$ (thin film) and $W_{31}Ta_{34}Cr_{5.0}V_{27}Hf_{3.0}$ (thick film) as a function of temperature were carried out. The chemical short-range order parameters (SRO) as a function of temperature as calculated from a 16000 atom simulation cells of the HEAs is plotted in Figures 6 and 7, respectively. The two compositions demonstrate different first order transitions and ODTT which reflect on the distribution of precipitates in the alloys at various temperatures. Among the considered pairs of atoms, the strongest attraction is observed for the Cr-Hf pair, which possesses the most negative SRO parameter at low temperatures for both HEAs. However, the temperatures at which the chemical order between Cr and Hf atoms vanishes are notably different, namely 1080 K for $W_{29.4}Ta_{42}Cr_{5.0}V_{16.1}Hf_{7.5}$ and 620 K for $W_{31}Ta_{34}Cr_{5.0}V_{27}Hf_{3.0}$ alloy (see Fig. 6). As a consequence, the Cr-Hf precipitate is visible in the simulation cell of $W_{31}Ta_{34}Cr_{5.0}V_{27}Hf_{3.0}$ alloy in Fig. 7a only at 300 K, whereas in the case of alloy with higher Hf concentration, it is observed in the three chosen simulation cells up to 1000 K (see Fig. 7b).

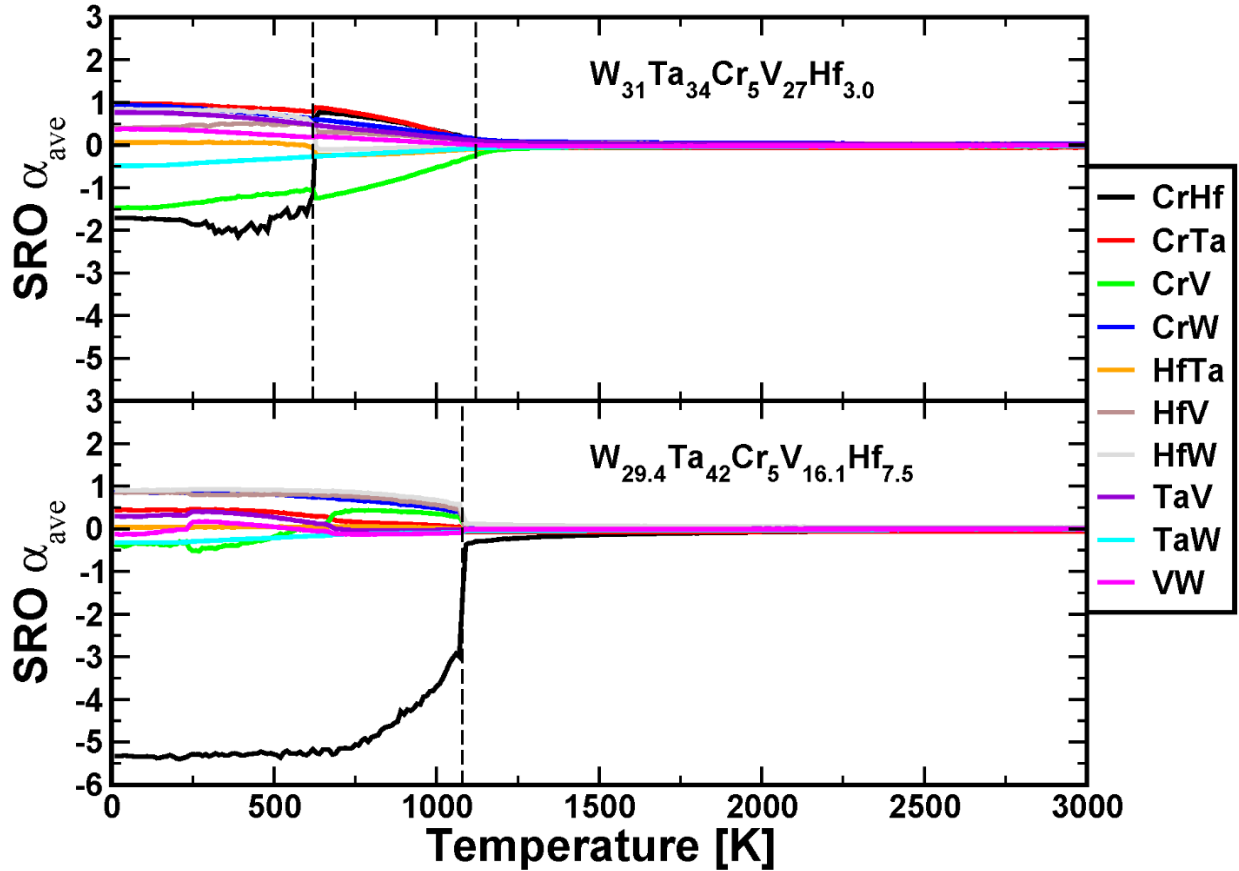

**Figure 6:** Chemical short-range ordering for different pairs of atoms as a function of temperature for both the $W_{29.4}Ta_{42}Cr_{5.0}V_{16.1}Hf_{7.5}$ (thin film used for the in-situ experiments) and $W_{31}Ta_{34}Cr_{5.0}V_{27}Hf_{3.0}$ (thick film used for mechanical properties and APT studies) alloys

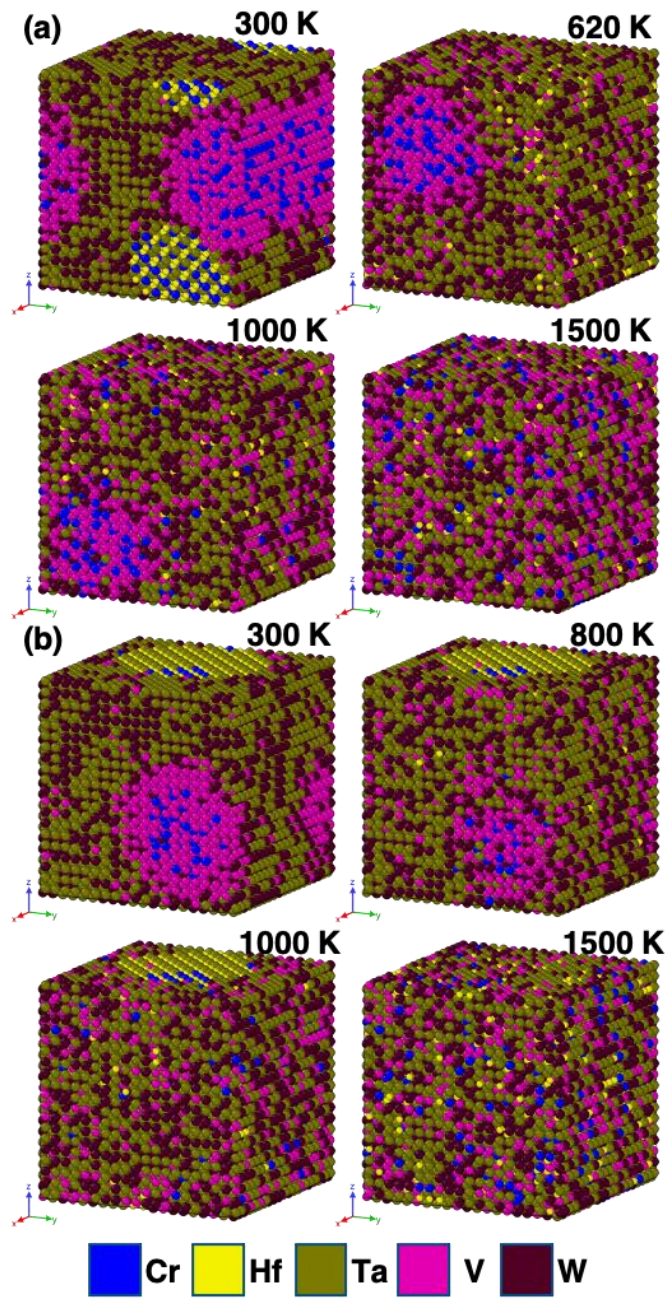

**Figure 7:** 16000 atom simulation cells of the (a) $W_{31}Ta_{34}Cr_5V_{27}Hf_{3.0}$ (thick film) and (b) $W_{29.4}Ta_{42.0}V_{16.1}Cr_{5.0}Hf_{7.5}$ (thin film) alloys as a function of temperature. Temperatures were chosen to reflect the results from Figure 6.

## 4. Discussion

### 4.1. Thermal stability

The synthesized quinary W-Ta-Cr-V-Hf RHEA possesses high thermal stability in terms of grain coarsening. Even though there was an increase in the grain size during He implantation, the material preserved its nano-crystallinity and showed only approximately 10 nm change in grain size after ~ 8.45% He implantation. The previously studied quaternary W-Ta-Cr-V RHEA demonstrated a similar thermal stability regarding grain morphology.[28] One of the properties that distinguishes this quinary W-Ta-Cr-V-Hf RHEA from the previously studied RHEAs is the grain refinement observed during dual beam irradiation. Grain refinement was observed before in other materials.[58] Two main suggested mechanisms are discussed in the literature: A) defect clusters produced during irradiation can migrate to sub-grain boundaries and form cell structures that lead to small grain formation[58] and B) cascades that are larger than the grain size can form a stacking fault across the grain breaking it into two separate crystalline structures (observed in FCC material).[59] Overlapping cascades that are smaller than grain size are also believed to potentially cause grain refinement.[58] The results here demonstrate that grain refinement only occurred when the heavy-ions are introduced. Even with 16 keV He ions, displacement should occur, and defects are then generated via Frenkel-pair production. However, grain refinement was not observed for the He only case, but rather grain growth occurred as shown in Figure 3. This suggests that such a grain refinement observed in the dual-beam irradiation case was caused by a cascade effect. The *in-situ* nature of the experiments performed in this work has allowed us to observe the grain refinement effect in real-time, albeit the division of some grains into sub-grains was observed at discrete time intervals since the fragmentation process was occurring at faster time scales than the frame rate used during the data recording. Figure 8 shows grain fragmentation that occurred during the dual beam irradiation. Molecular dynamics simulations have shown that high energy (50 keV) recoil events begin to divide in sub-cascades,[60,61] each with a diameter on the order of 10 nm. This upper limit on cascade size is near the initial grain-size of the W-Ta-Cr-V-Hf RHEA, allowing fragmentation to take place whereas in coarser-grained materials, a similar event would likely result in defect migration to boundaries. Under heavy ion irradiation, this effect was able to support and maintain the alloy's microstructural stability and even demonstrated grain refinement. As He implantation only causes grain growth and dual-beam irradiation leads to grain refinement, it can be expected that the single-beam irradiation results in further grain refinement. Therefore, if used in a nuclear environment where cascades are expected due to highly energetic neutrons, this material is expected to keep its structural integrity in terms of grain size and induced via grain refinement.

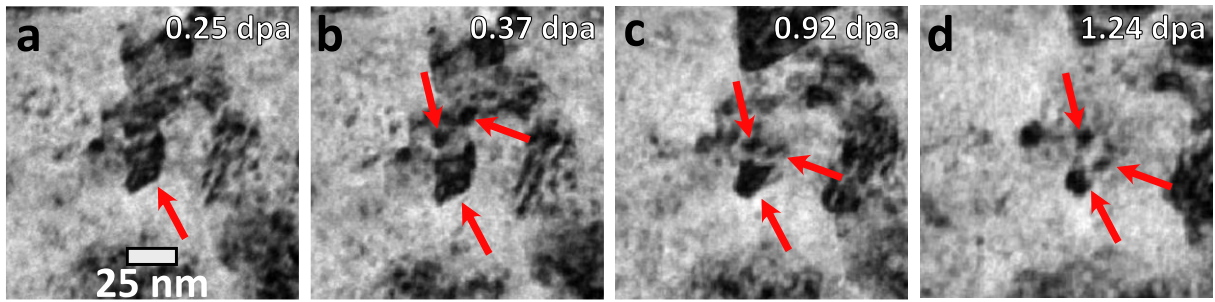

**Figure 8:** BF-TEM snapshots from the *in-situ* irradiation TEM video tracking the cascade-induced grain fragmentation during irradiation

Another significant improvement of this material system over the previous quaternary W-Ta-Cr-V RHEA, is its stability in terms of phase separation or segregation. The EDX demonstrated depletion of Hf from grain boundaries during irradiation which can be attributed to the Inverse Kirkendall effect (IKE)[62] or ballistic mixing during irradiation.[63] To demonstrate further the microstructural stability of this material, APT (Figure 9) was performed on thick films of the as-deposited sample and the irradiated *ex situ* single beam irradiated $W_{31}Ta_{34}Cr_5V_{27}Hf_3$ as it was extremely challenging to perform APT on the *in-situ* irradiated thin films (~100 nm). The APT results from the as-deposited sample demonstrated compositional striations that were not detected by EDX (which was also the case for the W-Ta-Cr-V RHEA).[28] The striations contained two main layers, a W-Ta layer and a Cr-V-Hf layer. The APT needle extracted from the irradiated sample contained two parts: deep unirradiated, but heated to 1073 K and an irradiated part at 1073 K. The compositional striations were shown to persist in the heated and un-irradiated part, but Hf and Cr have lost their correlation which is predicted by the SRO first order transition (Fig 6). The irradiated part showed homogenization of the elements and no precipitation. This is unlike the W-Ta-Cr-V RHEA where irradiation led to uniform and dense Cr-V precipitation.[28] This can be discussed based on the SRO as a function of temperature (Figure 6) and obtained through the modeling analysis. A negative SRO indicates ordering and a tendency to form intermetallic precipitates, while a positive SRO indicates tendency for phase separation. Homogenization of the elements occurs when all SRO values tend to zero which marks the ODTT of the alloy. The $W_{31}Ta_{34}Cr_5V_{27}Hf_{3.0}$ showed an ODTT of ~ 1120 K which roughly coincides with the irradiation temperature. Considering the irradiation-induced ballistic mixing and the corresponding atomic displacement and ordering,[63] homogenization can occur at lower temperatures. Therefore, we expect to see homogenized chemical distribution of the material after irradiation which was the case in this work. For the *in-situ* irradiated samples, TEM images and EDX showed no spherical precipitates and for that composition ($W_{29.4}Ta_{42}Cr_{5.0}V_{16.1}Hf_{7.5}$), the SRO and the corresponding atomic configurations (Figures 6&7) showed a slightly smaller (1080 K) ODTT (but rather different behavior prior to the ODTT) and, therefore, no precipitation is expected during the performed irradiation at 1173 K. It should be stated that the SRO and atomic configurations of the W-Ta-Cr-V RHEA predicted both density and size of the Cr-V precipitation.[28] We then conclude that: 1) no precipitation is expected in the studied compositions at temperatures close or over the ODTT, 2) changing composition can modify SRO and the material performance as a function of temperature and 3) a remarkable agreement between the experiments and modeling, thus allowing better understanding and prediction of the material's morphology as a function of temperature, which would constitute a new material design paradigm for developing RHEAs/HEAs for nuclear applications with different temperature requirements.

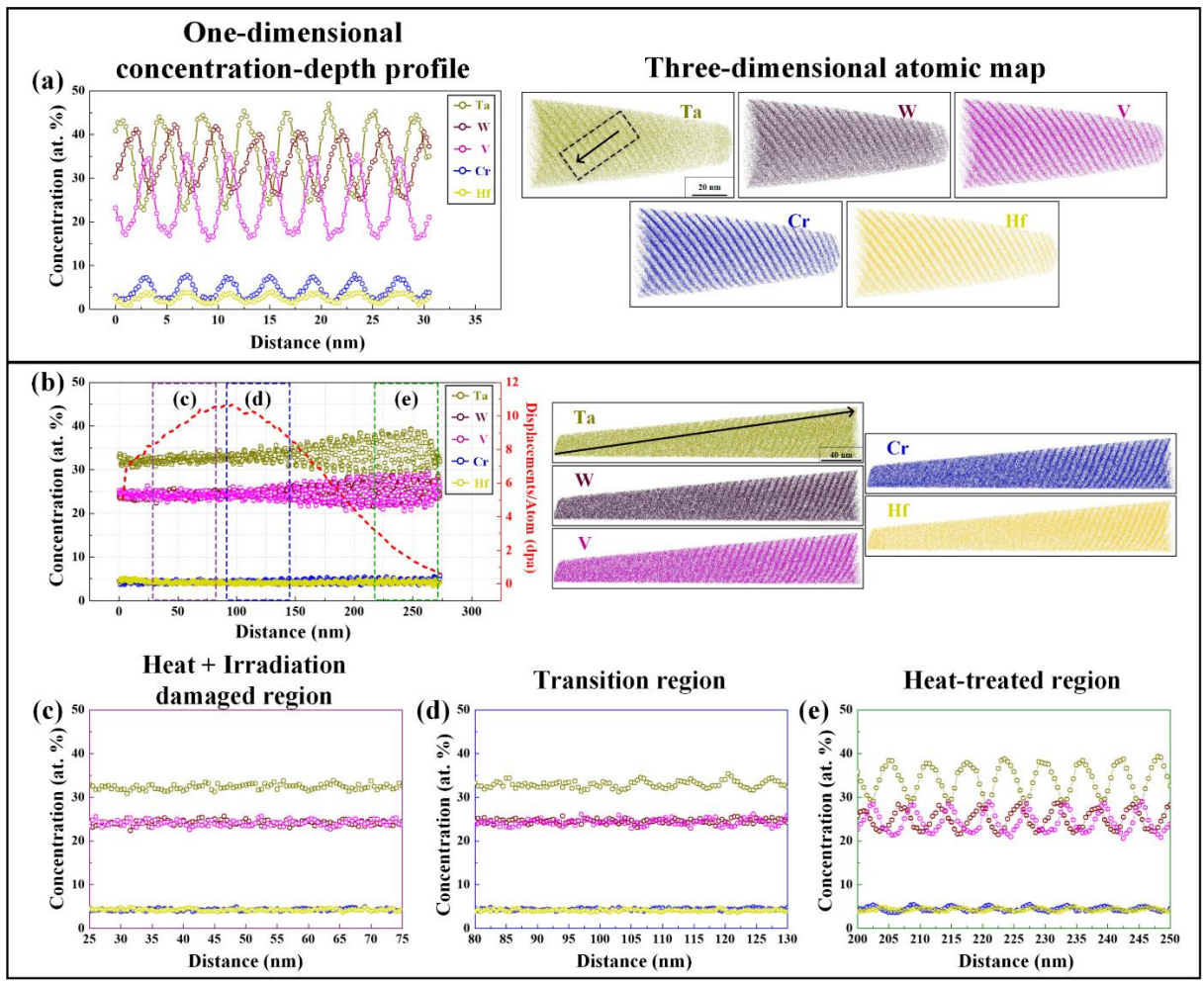

**Figure 9:** (a) APT atom maps of the as-deposited condition and the corresponding 1D concentration profile. (b) APT atom maps of the *ex-situ* heavy-ion irradiated thick film alloy ($W_{31}Ta_{34}Cr_{5.0}V_{27}Hf_3$) and the corresponding 1D concentration profile (left axis) the and dpa vs. depth profile (right axis). (c)-(e) 1-D chemical profiles showing the transition high dpa to low dpa regions.

### 4.2. Irradiation Response and Strength

The ion irradiation response of this alloy is characterized by: 1) no dislocation loop formation, 2) negligible change in overall volume due to bubble formation, 3) no preferential bubble formation on grain boundaries despite the high percentage of He implantation at 1173 K, 4) thermally stable grain-size (with 10 nm increase during He implantation up to 8.45%) with grain refinement under dual-beam ion irradiation, and 5) no precipitation when used at reactor relevant temperature (due to modest ODTT values). Pure W, for example, suffers with the formation of large facetted bubbles on the grain matrices and preferential bubble formation with larger sizes on the grain boundaries when implanted with the same He energy but at an even lower implantation value (6.3%).[64] At 1173 K, several He-vacancy complexes are mobile in W[65] and can coalesce to reach a change in volume of ~ 1.7% from the grain matrices contribution only. The grain boundaries, with much larger bubbles (~ 200 $nm^2$ compared to ~ 25 $nm^2$ in the grain matrices) had a larger contribution of ~7.4% change in volume.[64] Nanocrystalline W and ultrafine W-TiC alloys implanted with 2 keV $He^+$ at 1223 K to one order of magnitude lower fluence demonstrated 0.4% and 0.6% change in volume from the grain matrices respectively, and while preferential He bubble formation on the grain boundaries still occurred, its contribution for the quantification of change in volume was very challenging.[66] Under dual-beam irradiation, W demonstrated dislocation loop and dislocation network formation of high density and size (~ 100 $nm^2$ loop size and $0.5 \times 10^{-3}$ $nm^{-2}$ density) when the total dpa was only 0.25. The total change in volume was ~ 0.65 % (from grain matrices only).[56] The current alloy demonstrated no loop formation and the total change in volume was ~ 0.3 % after 8.45 dpa and higher percentage of implanted He. Additionally, no preferential large and facetted cavity formation at the grain boundaries was observed. It should be noted that possible surface effects are expected to diminish when the grain boundary to surface ratio approaches the value of 1.[67,68] In this HEA, the surface ratio is approximately 10. Furthermore, Surface effects in the HEA system is expected to be smaller than in pure W due to the rougher defect migration landscapes in HEAs[69], which can be applicable to some HEAs. [70]

The irradiation response to loop and cavity formation in W-based RHEAs was attributed to high interstitial-vacancy recombination,[28,71] which was also implied by Zhao who demonstrated via DFT calculations that vacancies in W-Ta-Cr-V RHEA possess smaller migration energies compared to pure W and overlapping interstitial and vacancy formation energies.[72] It is also shown by Zhao that most interstitial dumbbells are along the [110] direction,[72] which in BCC system, involves sequence of rotational and translational jumps as described by Schilling.[73] Others related the high irradiation resistance of HEAs to lattice distortion and sluggish diffusion effects[74] or the difficulty of defect clustering.[75] Using electronic structure calculations as implemented in VASP code, formation and migration energies of He were computed in the W-Ta-Cr-V system.[29] The He average migration energy was found to be 0.156 eV in the alloy compared to 0.06 to 0.081 eV in pure W and the He formation energy was ~2 times lower [76,77]. This rough energy landscape implies that He has a higher tendency in the alloy to quickly find a fairly stable site that can act as a bubble nuclei, and also to bind other slowly migrating He interstitial atoms. The clustering of He will slow them further down and a trend for smaller and lesser bubble nucleation will be enhanced, and therefore, uniform bubble distribution with no preferential bubble formation at the grain boundaries or a wide distribution of bubble sizes occur. The absence of preferential bubble formation at grain boundaries can also stem from high migration barrier of He-vacancy complexes, which still needs to be further investigated. Grain boundaries can also play a role in the irradiation resistance of the alloy acting as defect sinks. The contribution of the grain boundaries, however, compared to grain matrices in the alloy to the irradiation resistance has been studied comparing NC HEA and coarse grain HEA to NC-W and coarse grain W and it was found that the dominant factor is the grain matrix chemistry.[78]

### 4.3. Mechanical Strength

The radiation resistance of the alloy morphology has to be reflected on mechanical properties. Nanoindentation was performed for this purpose on one composition of this alloy system. It should be noted that different compositions of this alloy system can have different SRO parameters and different elemental segregation behavior (Figures 6 and 7), and therefore, are expected to have different mechanical properties. Here, we focus on one composition ($W_{31}Ta_{34}Cr_{5.0}V_{27}Hf_3$). The as-deposited sample had a hardness of 13.25 GPa. The increase in hardness after annealing was ~ 3 GPa. This is attributed to small voids in the films prior to annealing (which decreases the intrinsic strength) and possible segregation of elements at the grain boundaries after annealing (which increases the strength) at this temperature as revealed in Figure 2 (Hf segregation to the grain boundaries occurred in several tested different thin film compositions). After irradiation, the hardness further increased by 3.75 GPa. This HEA possesses high irradiation resistance and shows no dislocation loop formation, even under dual beam conditions (where He can bind to vacancies allowing interstitials to coalesce faster). Small defect clusters which are not visible in the TEM can affect the hardness. However, considering the dispersed barrier hardening (DBH) model[79] and other experimental work in W [80,81], the increase in hardness in this HEA cannot all be justified by the invisible defect clusters in TEM. Two other factors can affect the hardness of the irradiated HEA. First, some change in hardness could be a result of material homogenization which occurred due to the transition from ordered to disordered state, as discussed above, and which enhances lattice distortion.[82] Lee et al. [83] demonstrated improved mechanical strength in homogenized NbTaTiV HEA which were shown to stem from lattice distortion during deformation. Others have also suggested lattice distortion to be the dominant factor affecting the mechanical properties of HEAs. [84] Another factor that can cause a change in hardness in this HEA is the refinement of grain size when irradiated with heavy ion as demonstrated and illustrated earlier. Since the grain size in this HEA is very small (Figure 3) and hardness follows a power law with grain size, a small refinement of grain size could result in a large change in hardness. This change is higher when the grain size is very small and approaches the edge of the transition from the Hall-Petch to the Inverse Hall-Petch effect. [38] In Figure 4, the irradiated curve approached the annealed curve at high depths that exceed the irradiation depth by a factor of ~ 3 suggesting that the change in hardness (~ 1.6 GPa) at that depth is due to elemental segregation at the grain boundaries after annealing. The 3.75 GPA difference at 200 nm depth should include the elemental segregation (although it decreases after irradiation as shown in Figure 2), grain refinement due to irradiation, invisible clusters in the microscope, and the enhancement in lattice distortion due to homogenization of the elements. However, deconvolution of these effects is a complex process in a HEA system where several competing phenomena occur. However, this work will promote further studies on understanding this complex behavior of HEAs exposed to extreme irradiation conditions.

## 5. Conclusion

In summary, a new material design protocol, based on thermodynamic calculations and combined with experimental and simulation components, has been herein established for the development of promising and innovative irradiation-resistant RHEAs. Following this protocol, a new RHEA was designed and studied in terms of irradiation resistance, thermal stability and strength. The newly discovered HEA shows excellent irradiation resistance to dislocation loop and cavity formation under heavy ion irradiation and He implantation (even after 9% He implantation) with hardness values (in the annealed state) that are higher than the value in pure W. The irradiation resistance to loop and cavity formation under dual beam and He implantation is remarkably higher than other HEAs and pure metals studied in the literature. All results are studied and elucidated correlating the experimental results with modeling insight. We acknowledge that further works (*e.g.* response to neutron irradiation) are necessary to advance the technology readiness level (TRL) of this material and the effect of composition change in mechanical strength should be investigated. Ductility measurements should follow on coarse grain and bulk forms. Insights provided by the remarkable agreement between experimental and modeling results in this work can be used to design other novel alloys for different applications and constitute a new material design paradigm.